\documentclass[]{iucr}              
\usepackage{graphicx}
\usepackage{rotating}
\usepackage{longtable}
     \paperprodcode{a000000}      
     \paperref{xx9999}            
     \papertype{SC}               

     \paperlang{english}          
     \journalcode{J}              
     \journalyr{2007}
     \journaliss{1}
     \journalvol{63}
     \journalfirstpage{000}
     \journallastpage{000}
     \journalreceived{0 XXXXXXX 0000}
     \journalaccepted{0 XXXXXXX 0000}
     \journalonline{0 XXXXXXX 0000}

\begin{document}                  

\title{Improvements and considerations for size distribution retrieval from small-angle scattering data by Monte-Carlo methods}
\shorttitle{Monte-Carlo method improvements}

\author[a,b]{Brian R.}{Pauw}{brian@stack.nl}
\author[c]{Jan S.}{Pedersen}
\author[b]{Samuel}{Tardif}
\author[b]{Masaki}{Takata}
\author[a]{Bo B.}{Iversen}

\aff[a]{Center for Materials Crystallography, Department of Chemistry and iNANO, Aarhus University, DK-8000 Aarhus, Denmark}
\aff[b]{RIKEN SPring-8 Center, Sayo, Hyogo, 679-5148}
\aff[c]{Department of Chemistry and iNANO, Aarhus University, DK-8000 Aarhus, Denmark}

\shortauthor{Pauw, Iversen and Takata}

\maketitle

%

\begin{abstract}

Monte-Carlo (MC) methods, based on random updates and the trial-and-error principle, are well suited to retrieve particle size distributions from small-angle scattering patterns of dilute solutions of scatterers. The size sensitivity of size determination methods in relation to the range of scattering vectors covered by the data is discussed. Improvements are presented to existing MC methods in which the particle shape is assumed to be known. A discussion of the problems with the ambiguous convergence criteria of the MC methods are given and a convergence criterion is proposed, which also allows the determination of uncertainties on the determined size distributions.

\end{abstract}


\section{Introduction}

The search for generally applicable methods capable of determining structural parameters from small-angle scattering patterns for a broad range of samples has yielded several viable methods. For monodisperse systems consisting of identical particles, these methods attempt to find a free-form solution to the pair-distance distribution function $p(r)$, whereas for polydisperse systems the aim is to determine the distribution $P(R)$ in real space. In both cases, relevant transformation should  yield the observed scattering pattern \cite{Krauthauser-1996}. 

There are Indirect Transform Methods (ITM) based on regularization techniques which either impose that the solution is as smooth as possible \cite{Glatter-1977,Glatter-1979,Moore-1980,Svergun-1991,Pedersen-1994}, or Bayesian and maximum entropy ITM methods, which find a most likely solution using a Bayesian approach and entropy maximization, respectively \cite{Hansen-2000,Hansen-1991}. There are also methods available based on Titchmarsh transforms for determining size distributions \cite{Mulato-1996,Fedorova-1978}.

Another class of methods, such as the Structure Interference Method (SIM) \cite{Krauthauser-1996} and some Monte-Carlo (MC) methods \cite{Martelli-2002,DiNunzio-2004}, assume a particular shape and do not appear to require smoothness constraints. These only have a positivity constraint and have so far been limited to size distributions of sphere-shaped scatterers. These methods can be used to extract the particle size distribution function of systems of (dilute\footnote{The use of ``dilute'' means that the data should not be influenced by concentration effects.}) scatterers whose shape is known or assumed \cite{Krauthauser-1996,Martelli-2002,DiNunzio-2004}. The MC variant approaches the optimization by trial-and-error, whereas the SIM uses a conjugate gradient approach \cite{Krauthauser-1994}. Both are conceptually easier than the ITM or those based on Titchmarsh transforms, and provide stable and unique solutions \cite{Martelli-2002,DiNunzio-2004,Krauthauser-1996}. 

Upon implementation of one such method by \citeasnoun{Martelli-2002}, hereafter referred to as ``The Martelli method'', several noteworthy changes were made in the present work. We give a brief summary of the working principle, and highlight the differences from the Martelli method. Then, a general solution for detection limits is derived for particles in a polydisperse set. This aids the MC method as it allows for improved contribution scaling during the optimization procedure and indicates detectability limits in the final result. Lastly, a convergence criterion is defined for the MC method, allowing for the calculation of uncertainties in the resulting size distribution. This method is applied to scattering data obtained during the synthesis of AlOOH nanoparticles. 

\section{A brief overview of the implemented method}\label{sc:imp}

\subsection*{Step 1: Preparation of the procedure}
The initial guess of the scattering intensity is calculated for a uniform random distribution of a number of spheres $n_s$ anywhere between size bounds $0\leq R_{sph}\leq \frac{\pi}{q_{min}}$ (where $q_{min}$ is the smallest measured value of $q=\frac{4\pi}{\lambda}\sin(\theta)$, with $\lambda$ the wavelength of the radiation and $2\theta$ the scattering angle), using:
\begin{equation}\label{eq:main}
I_\mathrm{MC}(q)=\sum^{n_s}_{k=1} \mid F_\mathrm{sph,k}(qR_k) \mid^2 R_k^{(6-p_c)} p^\ast (R_k)
\end{equation}
Where $F_\mathrm{sph,k}(qR_k)$ is the Rayleigh form factor for a sphere, $R_k$ the radius for sphere $k$, and the pseudo-size distribution $p^\ast (R)$ is related to the number distribution $p(R)$ through $p^\ast (R)=R^{p_c} p(R)$. In other words, in this calculation (for reasons detailed in paragraph \ref{sc:obs}) the volume-squared scaling of each sphere contribution is partially compensated for by an exponent $p_c$, so that the scaling follows $R_\mathrm{sph}^{(6-p_c)}$, with $p_c$ typically between $2\leq p_c \leq 4$. Incidentally, setting $p_c=3$ makes $p^\ast (R)$ identical to the volume-weighted size distribution.

Subsequently, the MC intensity is brought in line with the measured intensity through optimization of the scaling factor and background level using a least-squares residual minimization procedure, minimizing the reduced chi-squared $\chi^2_r$ \cite{Pedersen-1997}:
\begin{equation}\label{eq:chisqr}
\chi^2_r=\frac{1}{N-M}\sum^N_{i=1}\left[\frac{I_\mathrm{meas}(q_i)-I_\mathrm{calc}(q_i)}{\sigma_i}\right]^2
\end{equation}
where
\begin{equation}\label{eq:Icalc}
I_\mathrm{calc}(q_i)=A\times I_\mathrm{MC}(q_i) + b
\end{equation}
and where $N$ is the number of data points, $M$ the number of degrees of freedom (unfortunately ill-defined in an MC model, and is set equal to two here for the intensity scaling parameter $A$ and background contribution parameter $b$), $I_\mathrm{meas}$ and $I_\mathrm{calc}$ the measured intensity and calculated model intensity, respectively, and $\sigma_i$ the estimated error on the data point (the estimation method for the error is detailed in paragraph \ref{sc:conv}).

\subsection*{Step 2: Optimization cycle}
The Monte-Carlo optimization cycle then begins, by picking a random sphere from the set of $n_s$ spheres, changing its radius to another random value within the bounds, recalculating the intensity of the entire set using equation \ref{eq:main} and reoptimizing the scaling factor and background level (eqn. \ref{eq:chisqr} and \ref{eq:Icalc}), and checking if this radius change improves the agreement between measured and MC intensity, i.e. if the change reduces the $\chi^2_r$-value. If it does, the change is accepted, otherwise rejected. A rejection-acceptance mechanism (occasionally accepting ``bad moves'') was found not to be necessary. 

This method differs from the Martelli method, in that the Martelli method continually attempts to add new sphere contributions to an ever growing set, leaving the prior established set contributions untouched. The adaptation presented here leaves the number of sphere contributions in the set unchanged, but repeatedly tries to change the radius of a random contribution in the set.

\subsection*{Step 3: Convergence and post-optimization procedures}
The optimization is stopped once the condition $\chi^2_r<1$ has been reached (c.f. paragraph \ref{sc:conv}). If convergence has not been reached within a certain number of steps (here set to 1 million) for a limited number of attempts, the pattern is considered unsuitable for fitting with this method \footnote{It is very rare that a pattern is not described at the first attempt but can be described upon repetition, but it does happen occasionally.}. For visualization and analysis purposes, the set of spheres can be distributed in a histogram (weighted to compensate for $p_c$). The whole MC procedure is then repeated several tens of times in order to obtain information on the mean and standard deviation of the histogram points. 

\section{Observability of isolated spheres in a polydisperse set}\label{sc:obs}

As the scattering intensity of particles scales proportional to their volume squared (radius to the sixth power for spheres), the scattered intensity of smaller particles in a polydisperse set is quickly drowned out by the disproportionally larger signal of larger particles. This effect, however, is partially compensated for by the different q-dependence of the scattering of the smaller particles. 

To investigate how large this compensatory effect is, we can define the ``maximum observability'' of a particle in a set as the maximum fractional contribution of that component to the total scattering pattern. I.e. the observability $Obs_{max,i}$ for component $i$ in a scattering pattern of $N$ independent contributions, measured within the $q$-range delimited by $q_\mathrm{min}\leq q_\mathrm{max}$ is defined as:
\begin{equation}\label{eq:mobs}
\mathrm{Obs}_{max,i}=\frac{I_i(q_m)}{ \sum_{j=1}^{N} I_j(q_m) }
\end{equation}
where the location $q_m$ is the value within the observed q-range where $\mathrm{Obs}_{max,i}$ is largest. For the MC method, both $I_i(q_m)$ and $I_j(q_m)$ are defined by equation \ref{eq:main}.

\begin{figure}
   \centering
   \includegraphics[angle=0, width=0.90\textwidth]{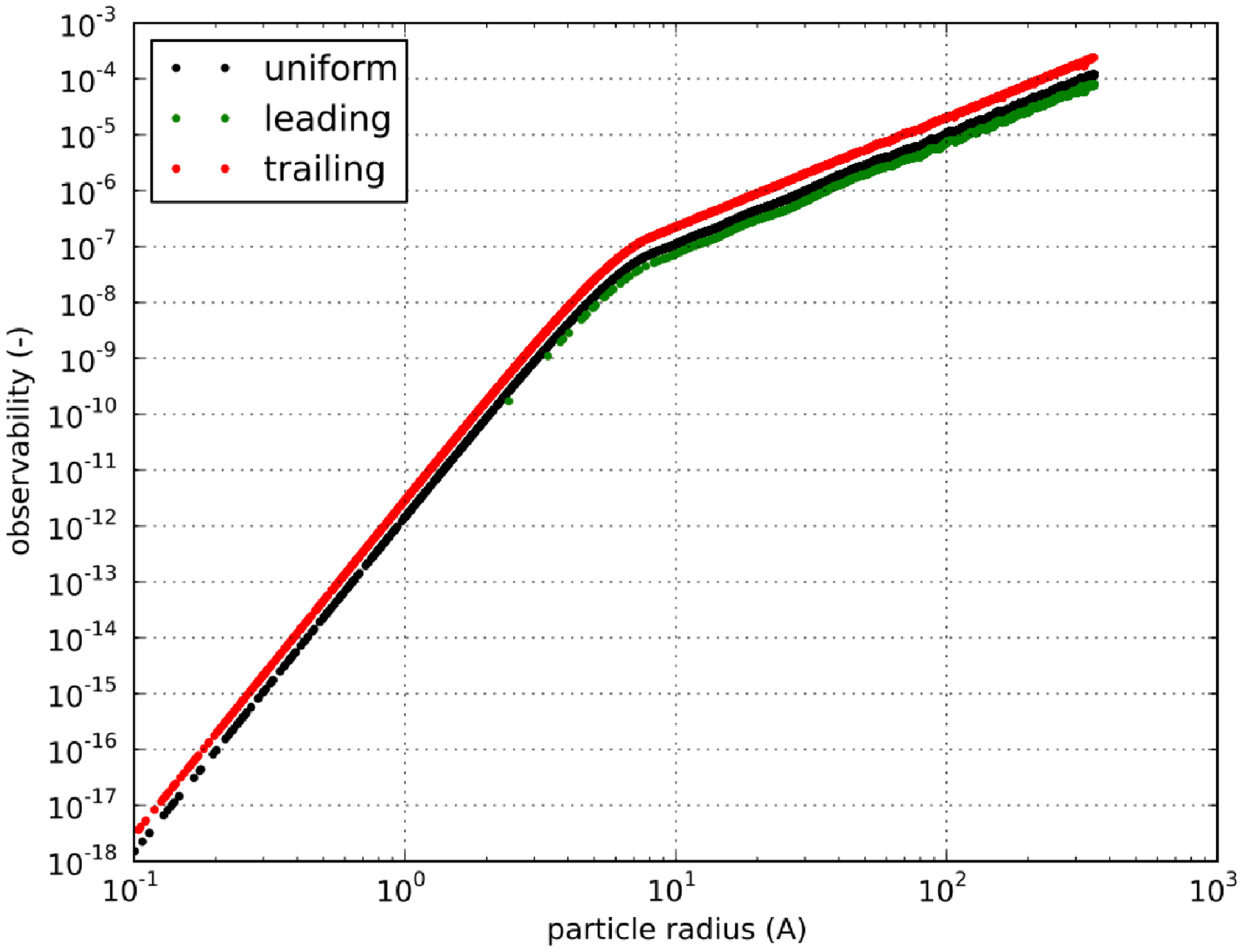} 
   \caption{Observability for three unimodal distributions of spheres (whose size distributions are shown in Figure \ref{fg:obsc}), within $0.01\leq q \leq 0.35$. A change in slope is observed at $R_\mathrm{lim}\approx \pi/q_{\mathrm{max}}$. $p_c$ is zero in the calculation.}
    \label{fg:obs}
\end{figure}

\begin{figure}
   \centering
   \includegraphics[angle=0, width=0.90\textwidth]{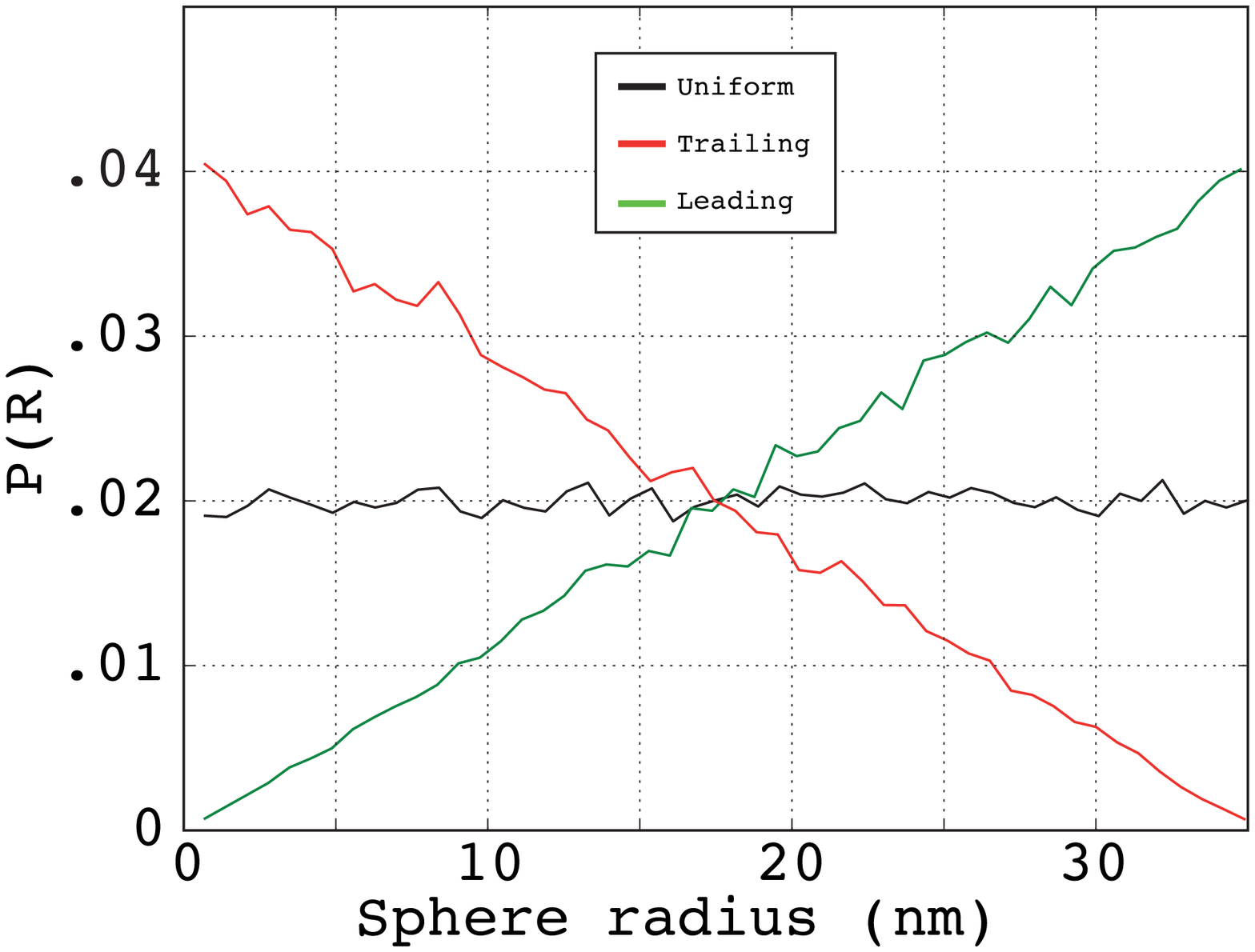} 
   \caption{Histograms (50 bins) of the distributions used for generating Figure \ref{fg:obs}}
    \label{fg:obsc}
\end{figure}

When plotting $\mathrm{Obs}_{max,i}$ for any size distribution of spherical particles, it is evident that the observability scales with the sphere radius squared (c.f. Figure \ref{fg:obs}) for particles with sizes larger than $R_\mathrm{lim}\approx \pi/q_{\mathrm{max}}$. Particles smaller than $R_\mathrm{lim}$ exhibit an observability scaling in line with the volume-squared intensity scaling (i.e. radius to the sixth power). The observability is shown for three unimodal distributions (one uniform and two triangular distributions of 50000 spheres) with particle radii between $0.1\leq R(\AA) \leq 350$. The modes of the triangular distributions are set to $0.1$ and $350$ for the ``trailing'' and ``leading'' triangular distributions, respectively (the distributions are shown in Figure \ref{fg:obsc}). Figure \ref{fg:obs} shows that the shape of the distribution has little visible effect on the shape of the observability. The absolute values of the observability are slightly dependent on the distribution, but more significantly dependent on the number of contributions (as can be inferred from equation \ref{eq:mobs} directly, but is not graphically shown).


The information on the observability can be used for three purposes. First and foremost, there is a clear indication of the limits of small-angle scattering for resolving the smaller sizes, and there is a link between the smallest measurable sizes and the maximum measured $q$ (as the observability scales with radius to the sixth power for particles with radii smaller than $\pi/q_\mathrm{max}$, their contribution vanishes rapidly). Note that it does not give information about the upper particle size detection limit, which is defined by the ability to distinguish the differences of scattering patterns of large particles.

Secondly, the knowledge can be used to calculate, for any model fitting solution, the minimum number of particles $n_{min,i}$ required to make a measurable impact on the total scattering pattern. This is done by calculating the inverse observability for the resulting distribution, and multiplying this with the overall measurement accuracy $\nu$ (usually 1\%).
\begin{equation}\label{eq:nobs}
n_{min,i}=\frac{ \nu \sum_{j=1}^{N} I_j(q_m) }{I_i(q_m)}
\end{equation}
Note that for the MC method presented here, in order to compensating for the discrepancy between the number of sphere contributions $n_s$ used and the number of bins $N$ they eventually end up in, we need to calculate:
\begin{equation}\label{eq:nobscomp}
n_{minMC,i}=n_{min,i}\frac{n_s}{N}
\end{equation}

In this way, any plots of size distributions can contain a line indicating a rough estimate for the minimum detectable number of particles (c.f. Figures \ref{fg:past} and \ref{fg:p}), which can prevent drawing erroneous conclusions from analysis artifacts. As stated above, the observability, and therefore these minimum required particle numbers, are directly dependent on the number of size divisions in the distribution.

Thirdly, the disproportionate contribution of larger spheres in numerical integrations over size distributions can be reduced by, instead of determining the size distribution function $p(R)$ in $I_\mathrm{MC}\propto\int \mid F_\mathrm{sph}(R) \mid^2 R^6 p(R) \mathrm{d}R$, to determine a pseudo-size distribution $p^\ast (R)=R^{p_c} p(R)$ in equation \ref{eq:main}. Upon determination of $p^\ast (R)$, the correct number-weighted distribution $p(R)$ can be retrieved through division of $p^\ast(R)$ by $R^p_c$. As indicated in paragraph \ref{sc:imp}, this compensatory power does not have to be equal to $2$, and can be tuned to further make the MC minimization more efficient.


\section{Data point weighting and convergence criterion}\label{sc:conv}

Estimates of the level of uncertainty on each measured data point (``errors'', $\sigma_i$ in equation \ref{eq:chisqr}) are invaluable to assessing the veracity of model fitting results (i.e. to determine whether the analysis provided a solution to within the uncertainty estimate). Additionally, its knowledge can help unlink the model fitting result from more arbitrary parameters such as the measured intensity integration bin width or the number of data points. By weighting of the goodness-of-fit parameter (used in the least-squares minimization function) by this error (c.f. Equation \ref{eq:chisqr}) uncertainties on the MC solution can be established.

These errors can be estimated to be \emph{at their very least} the counting-statistics-based Poisson error ($\sigma=\sqrt{I}$ where $I$ is the number of detected counts). Furthermore, if this estimate is exceeded by the sample standard error of the mean of the values contained in each individual integration bin (defined in equation \ref{eq:sse}), the sample standard error should be the preferred error estimate for that bin (as it can account for some detector irregularities). This sample standard error for each bin ($E_{\mathrm{bin}}$) is defined as:
\begin{equation}\label{eq:sse}
E_{\mathrm{bin}}=\sqrt{\frac{1}{N_p-1}\sum^{N_p}_{i=1}\left(I_i-I_{\mathrm{bin}}\right)^2}
\end{equation}
Where $I_{\mathrm{bin}}$ is the mean intensity in the bin, $I_i$ the intensity of datapoint $i$ in the bin and $N_p$ the total number of points in the bin.

Lastly, the absolute uncertainty of the measured intensity is commonly challenging to be below 1\%. Thus, 1\% (or however small the instrumental error is estimated to be) of the measured intensity should be preferred if it exceeds the other two estimates, These errors can then be used in equation \ref{eq:chisqr} to determine the goodness-of-fit parameter as acceptance-rejection criterion for a MC proposition.

The advantage of using these errors in the expression for $\chi^2_r$ is that if this parameter drops below one, the deviations between model- and measured intensities are \emph{on average} smaller than the statistical uncertainties. This thus provides a cut-off criterion for (for example) the MC method, allowing for the estimation of the mean and standard deviations of the final particle size histogram (shown in Figure \ref{fg:p}). Additionally, by using these errors in the expression for the goodness-of-fit parameter, the intensities are weighted by their relative errors in the fitting procedures and thus become less sensitive to arbitrary values such as bin widths or number of data points used in the fit.

Since the MC method does not provide us with an intensity at the same level as the measured intensity, and since there often is a constant background associated with small-angle scattering patterns for a variety of reasons \cite{Ruland-1971,Koberstein-1980}, these two parameters will have to be determined separately. Thus, after every MC proposed change, but before calculation of the goodness-of-fit, an intermediate least-squares minimization routine is applied to optimize the model intensity scaling and background parameters (eqn. \ref{eq:Icalc}). If required, the least-squares minimization method can be expanded to include more terms, at the cost of speed and stability. One reason for such an inclusion could be to include a power-law slope (with optional cut-off) to compensate for scattering from large structures or some inter-particle scattering effects \cite{Pedersen-1994,Beaucage-1995,Beaucage-1996}. 

\section{Uncertainties on the resulting distribution}

One common criticism of MC methods is the potential for ambiguity in the result, or the risk of over-fitting the data. While a rough estimate for the number of independently determinable radii (c.q. histogram bins) can be found using the sampling theorem (since $r_\mathrm{res}=\pi/q_\mathrm{max}$, $N\approx q_\mathrm{min}/q_\mathrm{max}$, assuming the largest measurable dimension is identical to the measurement limit \cite{Hansen-1991,Moore-1980,Taupin-1982}), it also has to be dependent on the uncertainty of the underlying dataset. An alternative practical way of investigating the result validity is to determine the errors (standard deviation in this case) on the result.

These can be obtained with the present MC method, by performing the MC fit to the same data several tens of times, and for each time optimizing until $\chi^2_r<1$ has been achieved. By distributing the results in histograms with a fixed array of bins, the mean value and standard deviation for each bin can be determined. Naturally, the relative standard deviation is a function of the bin size, so that more numerous narrower bins will have a larger standard deviation than fewer wider bins (within reason). Additionally, the level of the observability (equation \ref{eq:mobs}) is dependent on the number of bins, leading to a trade-off between minimum observable number of particles $n_{minMC,i}$ and the number of bins. For the most common equidistant histogramming, however, it is up to the user to determine the best suited number of size histogram bins (or to use a value close to the sampling-theorem-derived value).

The procedure then provides the user with a clear overview of the uncertainties and detection limits attached to the determination of polydispersity from small-angle scattering patterns, which can be used for further extraction of meaningful numbers from the resulting distribution. 

\section{Experimental}

\subsection{Synthesis}

Boehmite (AlOOH) particles were synthesized in-situ using an automated and modified version of a high-pressure high-temperature reactor \cite{Becker-2010}. The sapphire capillary in which the reaction takes place has an inner diameter of 1.0 mm and an outer diameter of 1.57 mm. The particles were synthesized from a solution of 0.5M Al(NO$_3$)$_3$ precursor in water. The start of the reaction was considered to be the moment at which the pressurized solution (maintained at a pressure of 250 bar) is heated to its reaction temperature of 275 degrees centigrade. The measurement used in this paper was obtained 1700 seconds from the start of the reaction. Further details and results will be presented in a forthcoming paper.

\subsection{Beamline details}

Synchrotron SAXS experiments were performed at the BL45XU beam line of the SPring-8 synchrotron in Japan. The beam was collimated to a 0.4 by 0.2 mm beam (horizontal by vertical, respectively), with photons with a wavelength of $0.09$ nm. The sample-to-detector distance was 2.59 meter. The scattering patterns were recorded on a Pilatus 300k detector whose total surface area covers 33.5 by 254 mm, consisting of 195 by 1475 pixels measuring 0.172 by 0.172 mm. Transmission values were determined using in-line ionization chambers. The polarization factor was assumed to be 0.95. The measurements were collected at a rate of 1 Hz. 

\subsection{Data correction}

The data were corrected for background (water at 275 degrees centigrade and 250 bar), incoming flux, measurement time, transmission factor, polarization factor, spherical correction factor and calibrated to absolute units using a Glassy carbon sample from series H, supplied by Dr. Ilavsky from APS \cite{Zhang-2009}. Statistics were calculated according to the procedure outlined in paragraph \ref{sc:conv}, with the minimum possible error set to 1\% of the measured intensity. 

\section{Results and discussion}

\begin{figure}
   \centering
   \includegraphics[angle=0, width=1\textwidth]{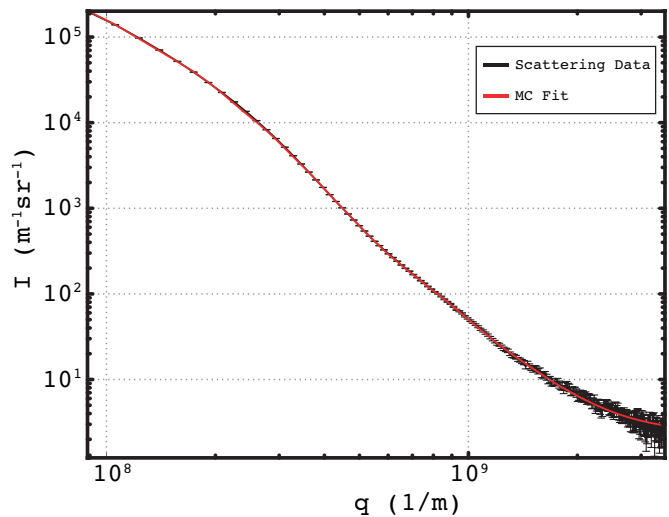} 
   \caption{Data (black) and MC fit (red, using 1000 spheres) for one one second measurement in a time series of AlOOH nano particles in aqueous solution. The MC fit is at convergence.}
    \label{fg:fit}
\end{figure}

\begin{figure}
   \centering
   \includegraphics[angle=0, width=1\textwidth]{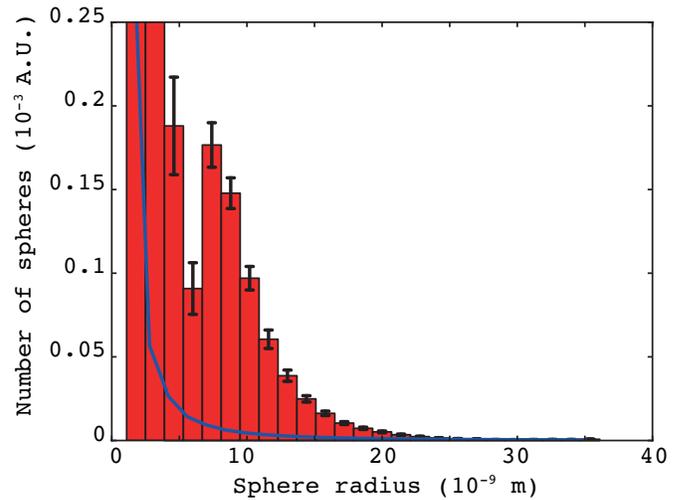} 
   \caption{Size distribution $P(R)$, standard errors and minimum observable number of particles obtained by transforming the distribution $P^\ast(R)$ shown in Figure \ref{fg:past}. Shown with limited vertical axis range for clarity.}
    \label{fg:p}
\end{figure}

\begin{figure}
   \centering
   \includegraphics[angle=0, width=1\textwidth]{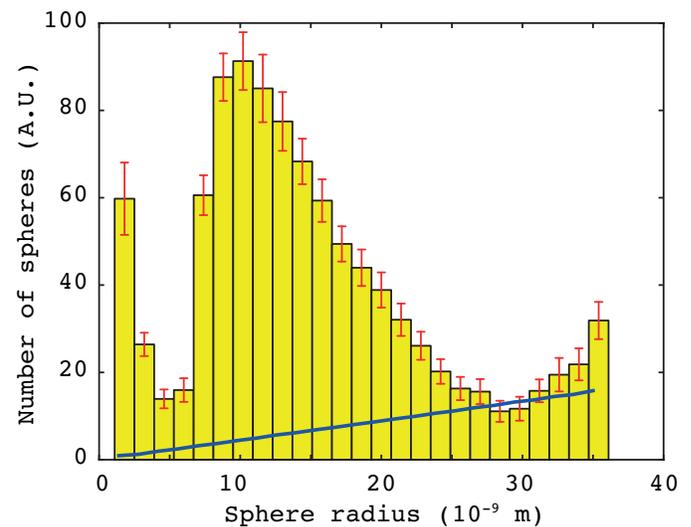} 
   \caption{Pseudo-size distribution $P^\ast(R)$ with $p_c=3$ as used for the MC fit shown in Figure \ref{fg:fit}. Error bars indicate sample standard deviation over 100 repetitions. Minimum observable number of spheres $n_{minMC,i}$ shown as blue line.}
    \label{fg:past}
\end{figure}

The collected dataset, its error and the MC fit are shown in Figure \ref{fg:fit}, where the MC fit intensity is the average of 100 repetitions of the MC procedure. While a single run also delivers a model intensity to within the determined error (on average), the mean intensity is shown here as it matches the mean of the size distributions shown in Figures \ref{fg:past} and \ref{fg:p}. In Figure \ref{fg:past}, the pseudo-size distribution is shown, as determined from the MC procedure using $p_c=3$. The error bars indicate the standard deviation of the histogrammed value for each of 100 repetitions. This figure also contains the line indicating the minimum number of visible particles required $n_{minMC,i}$ (which here is proportional to the radius due to the choice of $p_c$). 

Transformed to a number size distribution the histogram becomes that shown in Figure \ref{fg:p}. It is immediately clear that the larger particles are only present in the solution in minuscule amounts, and that there is a dip in the number of particles with a radius around 5 nanometer, with many particles slightly smaller and larger than that size. 

\begin{figure}
   \centering
   \includegraphics[angle=0, width=1\textwidth]{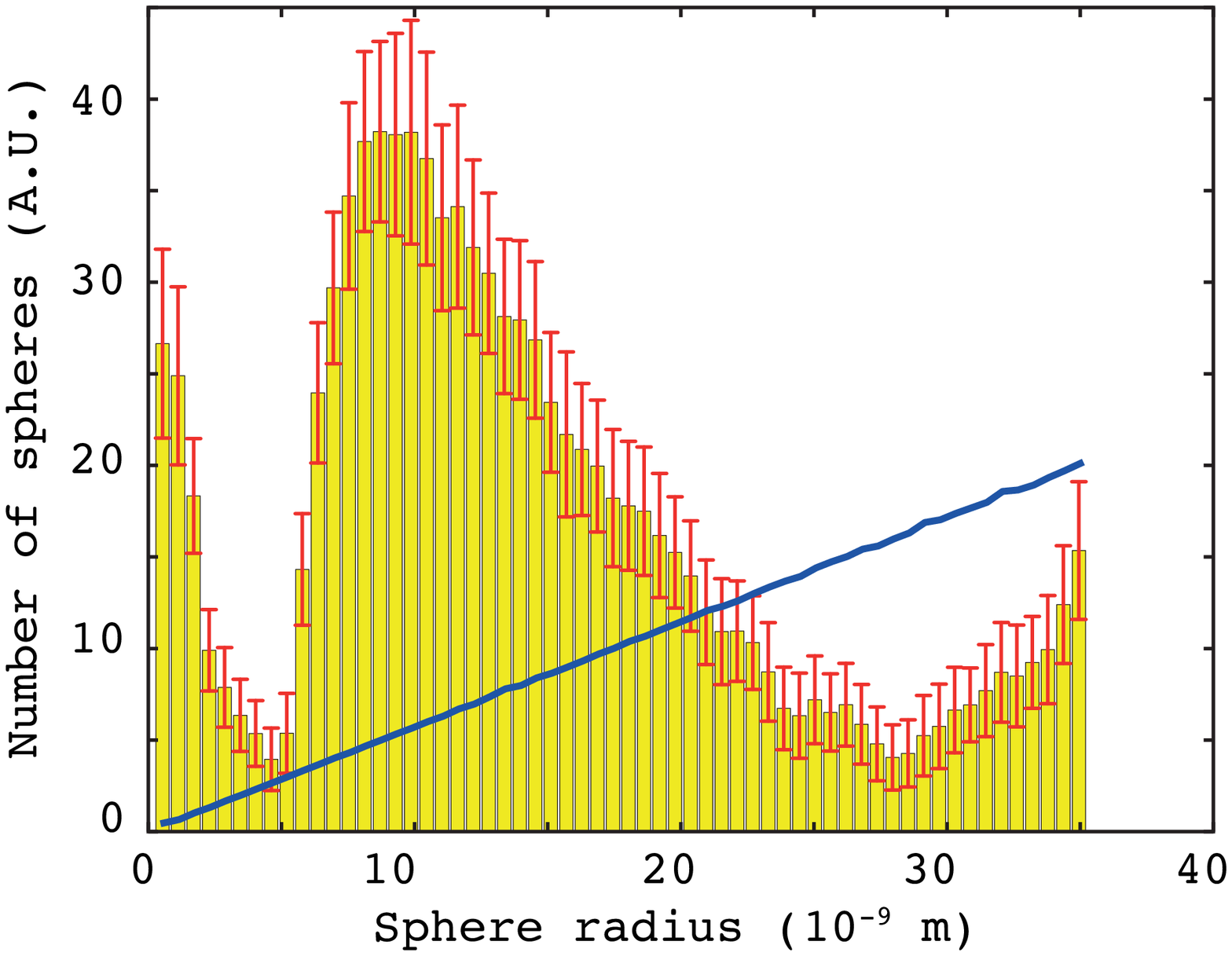} 
   \caption{The MC-determined $P^\ast(R)$ of Figure \ref{fg:past} histogrammed in 60 bins. Minimum observable number of spheres $n_{minMC,i}$ shown as blue line.}
    \label{fg:past60}
\end{figure}

\begin{figure}
   \centering
   \includegraphics[angle=0, width=1\textwidth]{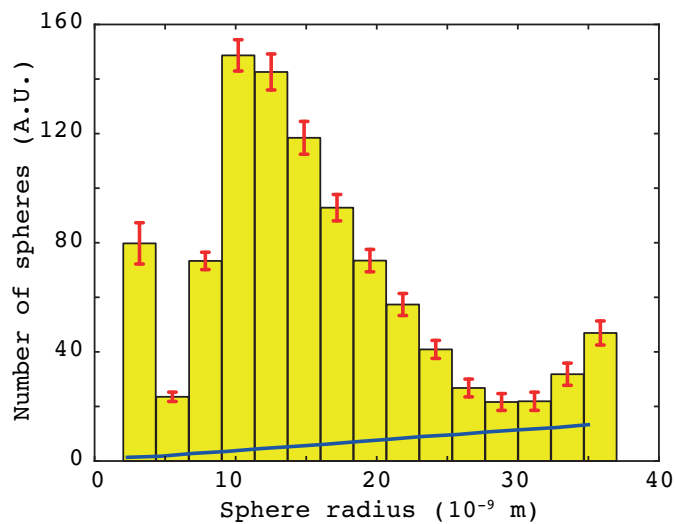} 
   \caption{The MC-determined $P^\ast(R)$ of Figure \ref{fg:past} histogrammed in 15 bins. Minimum observable number of spheres $n_{minMC,i}$ shown as blue line.}
    \label{fg:past15}
\end{figure}

In this example, the number of histogram bins is 25, but one can choose more or fewer bins. The effect of this is shown in the pseudo-size distributions in Figures \ref{fg:past60} and \ref{fg:past15} for 60 and 15 histogram bins, respectively. This clearly shows the relation between the standard deviation in the bins and the minimum observable number of particles. If the number of histogram bins is high, the uncertainty for each value is equally large, \emph{and} the minimum number of particles of each bin size required to make a measurable impact on the scattering pattern increases. If the number of bins is reduced, both the uncertainty and $n_{minMC,i}$ reduces, at the cost of detail. For further (and more involved) improvement in information content retrieval, the bins in Figure \ref{fg:past} that fall below the $n_{minMC,i}$ line could be combined (which would render them ``observable''), and the bins above this line could be further subdivided for improved information extraction. 


\section{One more thing...}

All the above results were obtained assuming that the scatterers are spherical in shape. This does not have to be the correct particle shape for the method to arrive at a solution. As mentioned before, the size distribution and shape cannot be uniquely separated from scattering patterns (which has been tested for simulated isotropic scattering patterns from polydisperse sets of prolate and oblate ellipsoids). The solution from the MC method, then, shows the user what the size distribution would be \emph{if} the scattering pattern originated from spherical particles. 

If the shape of the scatterers is known from other investigations such as electron microscopy, and deviates from a spherical shape, this information can be used to obtain the correct size distribution for that particular shape \cite{Pedersen-1996}. This can be done by either adjusting the particular scattering function in the MC method, or by analysis (or rather deconvolution) of the correlation function $\gamma (r)$ which can be calculated from the sphere-based MC method result \cite{Feigin-1987}. However, it should be kept in mind that only one length distribution can be uniquely determined due to the limited dimensionality (information content) of the isotropic scattering data.

\section{Conclusions}

Discussed in this paper are modifications to the Martelli MC method, the general veracity of the result, and the application of it to a SAXS measurement. It is shown that by using the methodology described in this paper, a particle size distribution can be retrieved from a scattering pattern, uncertainties can be estimated for the particle size distribution, and the minimum number of particles indicated for each size required to make a measurable impact on the scattering pattern can be indicated for each size. 

The MC code is available for inspection, improvements and application and will be freely supplied by the author upon request.

\section{Acknowledgements}

SPring-8, RIKEN and Dr. Hikima are acknowledged for their beamtime and support. Dr. Becker and Dr. Eltzholtz have been instrumental in the success of the measurements. 

\referencelist[/Users/brian/Documents/bibliography] 

\end{document}